\documentstyle[aps,preprint]{revtex}
\begin{document}

\begin{center}
{\large\bf Physics of Fashion Fluctuations}
\vspace{2.5cm}

{\large R.\ Donangelo$^a$, A.\ Hansen$^{b,c}$,
K.\ Sneppen$^b$ and S.\ R.\ Souza$^a$}
\vspace{1.5cm}

$^a$ Instituto de F\'\i sica, Universidade Federal do Rio de Janeiro,\\
C.P. 68528, 21945-970 Rio de Janeiro, Brazil.
\vspace{0.5cm}

$^b$ Nordita, Blegdamsvej 17, DK--2100 Copenhagen, Denmark.
\vspace{0.5cm}

$^c$ Institutt for fysikk, Norges Teknisk-Naturvitenskapelige Universitet,
N--7491 Trondheim, Norway.
\vspace{0.5cm}

\end{center}
\vspace{2.5cm}

\begin{small}
\noindent
Abstract: We consider a market where many agents trade many different types
of products with each other.  We model development of collective modes
in this market, and quantify
these by fluctuations that scale with time with a Hurst exponent of about 0.7.
We demonstrate that individual products in the model occationally
become globally accepted means of exchange, and simultaneously become
very actively traded. Thus collective features
similar to money spontaneously emerge, without any a priori reason.
\end{small}

\vspace{2.0cm}
\vfill\eject

One of the most important interactions among humans is the exchange of
goods and services.
Trade has a number of benefits. Besides strengthening social contacts,
it directly allows individuals to specialize while still benefitting
from a large diversity of products.
Another important benefit is risk minimization. Even in primitive markets 
it is beneficial for agents to have access to a diversity of goods,
thereby minimizing their dependence on any particular one.

In modern markets risk minimization is similarly obtained by
diversification of stock portfolios. 
However, diversification may be orthogonal to maximizing short
term profit, and the frustration between these two goals strongly
influences the behaviour of investors.

In the present paper we discuss a recently proposed minimal model
for a market where agents attempt both to obtain the goods they
need, by maintaining a stock of different products, and, at the 
same time, obtain what other agents want \cite{DS00}. 
In other terms, they try to fulfill both their ``need" and
their ``greed".
The model was inspired by Yasutomi's work \cite{Yasutomi} on
possible emergence of certain goods as ``money" in the sense that 
they become fashionable, and therefore desirable, in a market where
only barter trade takes place \cite{Menger}.  

The minimal model for the market we propose consists of $N_{ag}$
agents and $N_{pr}$ different products.
Initially we give $N_{unit}$ units of the products to each agent.
The number $N_{unit}$ is fixed, but the products are chosen at
random, so the individuals are not in exactly the same situation.
At each timestep we select two agents at random.  These will
attepmt a trade between themselves.
The trade starts by comparing the list of goods that each agent
lacks and therefore would like to get from the other agent in
exchange for goods it has in stock.

We first consider the simple ``need" based exchange procedure:
when each of the agents has products that the other needs, then
one of these products, chosen at random, is exchanged.

In case such a ``need" based exchange is not possible we consider
the ``greed" exchange procedure: one or both of the agents
accept goods which they do not lack, but consider useful for
future exchanges.

In order to determine the usefulness of a product, each agent $i$
keeps a record of the last requests for goods he received in
encounters with other agents. 
This memory is finite, having a length of $N_{mem} (i)$ positions, 
each of which registers a product that was requested.
Different agents may in principle have different memory length,
although we typically will set all to be equal, i.e.,
$N_{mem} (i) = N_{mem}$ for all $i$.
As the memory gets filled, the record of old transactions is lost.

The demand for a product is measured by each agent through the number
of times that product appears on its memory list.
Agents will then accept products they already have in stock with a
probability based on its memory record.
The chance that agent $i$ accepts a good $j$ is taken to be proportional
to the number of times $T_{ij}$ that good $j$ appears on the memory
list of agent $i$:

\begin{equation}
p_{ij} \;=\; \frac{T_{ij}}{N_{mem}(i)}\;,
\end{equation}

\noindent 
where we have used the fact that $\sum_j T_{ij} = N_{mem} (i)$.
These two exchange mechanisms define our model.

In a previous paper \cite{DS00} we have demonstrated that after a small
number of encounters per agent, the initial needs of all agents become
locally minimized and most trades are based on the ``greed" procedure.

As mentioned above, we typically we assign the same memory to all agents.
However, as an initial lesson, we would like to demonstrate that memory
helps agents in their trading.
To do this, we consider a system where $N_{ag}=200$ agents trade $N_{pr}=200$
different types of products, and where each agent is given $N_{unit}=400$
units of products.
To illustrate the effect of memory, we have performed two different
calculations, in which we arbitrarily assigned to each agent $i$ either 
a memory $N_{mem}(i)=i$ or $N_{mem}(i)=5i$.

Fig.~1 shows the time averaged trading probability of agents as function of 
their memory.
We notice that, in general, trading activity increases with memory size.
Thus, a long memory provides the agent with a better perception of the 
overall needs in the market, which leads them to value products others
agents also have in high demand.
One should note, however, that agents with very short memories also do
well, which probably means that an agent that trades practically at random
may do better than agents which know less than the typical trader. 
We should stress that what agents are optimizing here is their trading 
activity, and not their individual wealth.
Therefore this lesson may apply to agents that work for a commision on
transactions between real traders.

For the remaining of this paper we consider that all agents have the same
memory size, $N_{mem}(i)=N_{mem}$, and use the model to examine how different
agents allocate their stocks, and how particular products change from being 
fashions to being forgotten.

We now study some specific properties of the model. 
After the initial reshuffling of goods, during which, as we have mentioned,
mostly basic needs are fulfilled, each product will begin a slow development
towards an absorbing state, i.e., a state which will develop no further 
when reached.
Each product reaches this absorbing state when all agents have the product, 
and, at the same time, none of the agents remembers that any other agents 
needs the product, i.e., the product has disappeared from the memory
of all agents.
In this case all activity on that product disappears.

In Fig.~2 we show the number of active products as function of number of 
trades per agent (which is a measure of elapsed time) for a system with
size parameters
$N_{ag}=50$, $N_{pr}=50$, $N_{mem}=100$ and $N_{unit}=100$.
The upper curves in fig.~2 indicate that the decay towards the absorbing 
state is avoided through the inclusion of production/consumption processes.
Production/consumption is here defined as a probability $p$ that, at a given
time, an agent consumes a random product in his stock, and produces another 
he either needs or considers valuable. 
The rate of production $p$ in the two cases shown are one per 100 trades 
($p=0.01$) and one per 1000 trades ($p=0.001$) in the system.

When noticing the tendency of a few products to dominate the market, 
discussed in Ref.~\cite{DS00}, one would similarly observe that production 
limits this tendency. This is easily understood, because, if products are 
easily produced, they do not become needed and therefore valuable. 
Furthermore, we have checked that, if different products are produced with 
rates which are different a priori, it is the slowly produced products 
that become valuable.

The freezing of products into the absorbing state, implies that demand tends 
to concentrate on just a few products, i.e., it emerges a subset of 
goods that are globally accepted for trade.
During the slow transient towards the frozen state the total demand for a 
product $j$

\begin{equation}
D(j) \;=\; \sum_i T_{ij}\;,
\end{equation}

\noindent
fluctuates as a function of the number of trades $t$ as 

\begin{equation}
\Delta D(j) \propto (\Delta t)^H\;,
\end{equation}

\noindent
i.e., as a fractional Brownian walk with Hurst exponent
$H$ \cite{Hurst,f88}. 
We have found that, in the case of no production/consumption processes, 
for a large range of system sizes, $H=0.7\pm0.05$ \cite{DS00}.

In the case of finite production we have also found that fluctuations
in demand also exhibit fractional Brownian walk properties in the
statistically stationary state.
This is illustrated in Fig.~3 a and b where we show, for a given product,
the time series of its demand and its fluctuation statistics, respectively.
The calculations were done for the same system parameters as in Fig.~2,
and with production/consumption rate $p=0.001$.

We have thus seen that if we quantify the value of products through their
demand, then value becomes concentrated on a few products, and exhibits
persistent fluctuations ({\it i.e.} fluctuations with Hurst exponent $H>0.5$). 
Such basic features are commonly found in goods that are employed as
means of exchange, that is, serve as money. 
In order to quantify this common view of the what is valuable, we define
the monetary value of a good $j$ as the number of agents which consider
that good as the most wanted in the system,

\begin{equation}
M(j) = \sum_i \prod_{j'\ne j} S(T_{ij}-T_{ij'})\;,
\end{equation}

\noindent
where $S(x)$ is the step function, $S(x\ge 0)=1, S(x<0)=0$.

In Fig.~4 we examine the monetary value as function of demand for the
products in the system considered in Fig.~3.
We observe a clearly nonlinear relationship between monetary value and
demand, which reflects the extremal nature of $M$. 
We also notice that, when demand exceeds a certain critical value, the
product becomes the most globally accepted means of exchange, or, in
more popular terms it becomes money. 
As a consequence of this nonlinear relationship, the short time
fluctuations of monetary value of goods exhibit pronounced fat tails,
which gradually become Gaussian when considered over longer time
intervals~\cite{DS00}.

The increased demand for a product is closely linked to an increased 
heterogeneity of the distribution of products between agents. 
As a product becomes slightly more needed than the rest, agents start
demanding it, and accepting it even without need because they sense
other agents needing it.
As a result of the heavy trading in that good, and the resulting
fluctuations in its distribution throughout the system, the good may
be accumulated by a few agents, which further amplifies its need.
In the lower curves of Fig.~4 we show that demand for a product
is proportional to the number of agents that lack it.
The two lower curves show respectively the instantaneous number of agents
without the product and the number of agents that persistently over 
a time interval $\Delta t=2$ do not have the product.
This shows that although the most valuable product is 
unevenly distributed, the ownership will fluctuate on a fast timescale.

Finally note that the same fluctuations that make a good the most traded may also
distribute it evenly among agents. If the distribution lasts long enough for
other products to replace it in the memory of the agents in the system,
another product will become the most fashionable. 
Thus, the concept of money, and its stability, may be directly linked to
the fact that economic agents have memory.  

For random appearance of monetary systems, consult the case studies
of Yasutomi~\cite{Yasutomi}.
For quantifications of persistent fluctuations of value
in stock and monetary markets, see e.g., Ref.\ \cite{Evertz}. 
For discussions of other models which exhibit anomalous scaling, see also 
\cite{DS00,RHSS,BPS} whereas a detailed discussion of scaling phenomena
in economy may be found in the excellent review of Farmer \cite{Farmer}.
\vspace{0.5cm}

In summary, we have constructed a simple model for emergence of fashions ---
goods that become popular not due to any intrisic value, but simply because
``everybody wants it" --- in 
markets where people trade goods in order to fulfill the mutually frustrating 
demands of need and greed.
This model shows spontaneous emergence of random products as money.
The limitation of the model as an economic model is mainly due to the fact that
it does not incorporate profit, but only works with perceptions of value. 
Therefore a detailed comparison with financial markets is not attempted. 
Nevertheless the model supports collectively driven fluctuations characterized
by a Hurst exponent of $\approx 0.7$, and its sets a frame where one may study 
how production/consumption influences development and collapse of value.

\vspace{0.5cm}

A.H.\ thanks Nordita and the Niels Bohr Institute for hospitality and support 
during his sabbatical stay in Copenhagen during which this work was performed.

\vspace{0.5cm}
{\bf Figure Captions}

\begin{itemize}

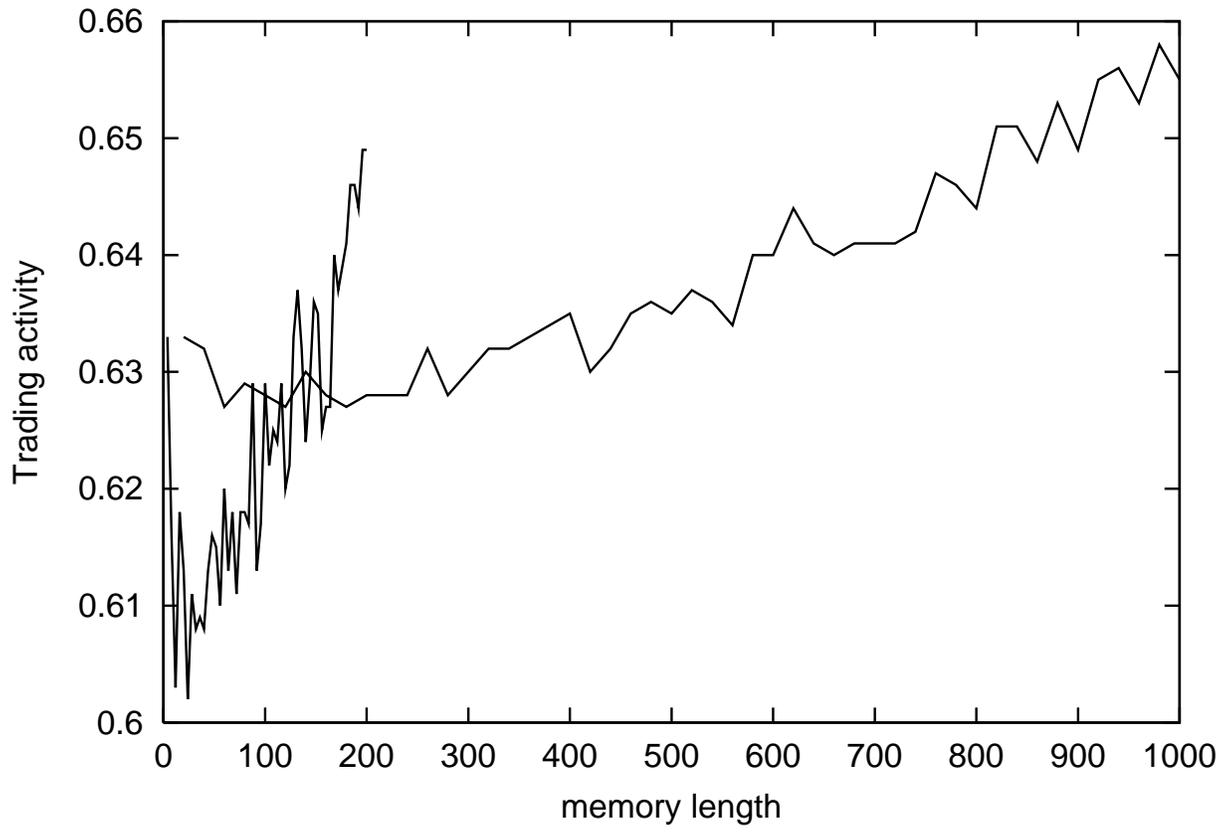
\item Fig. 1. Agent trading activity as function of length
of its memory in a market of 200 agents endowed with memories 
of sizes 1,2,...,200, or 5,10,...,1000. See text for more
details.

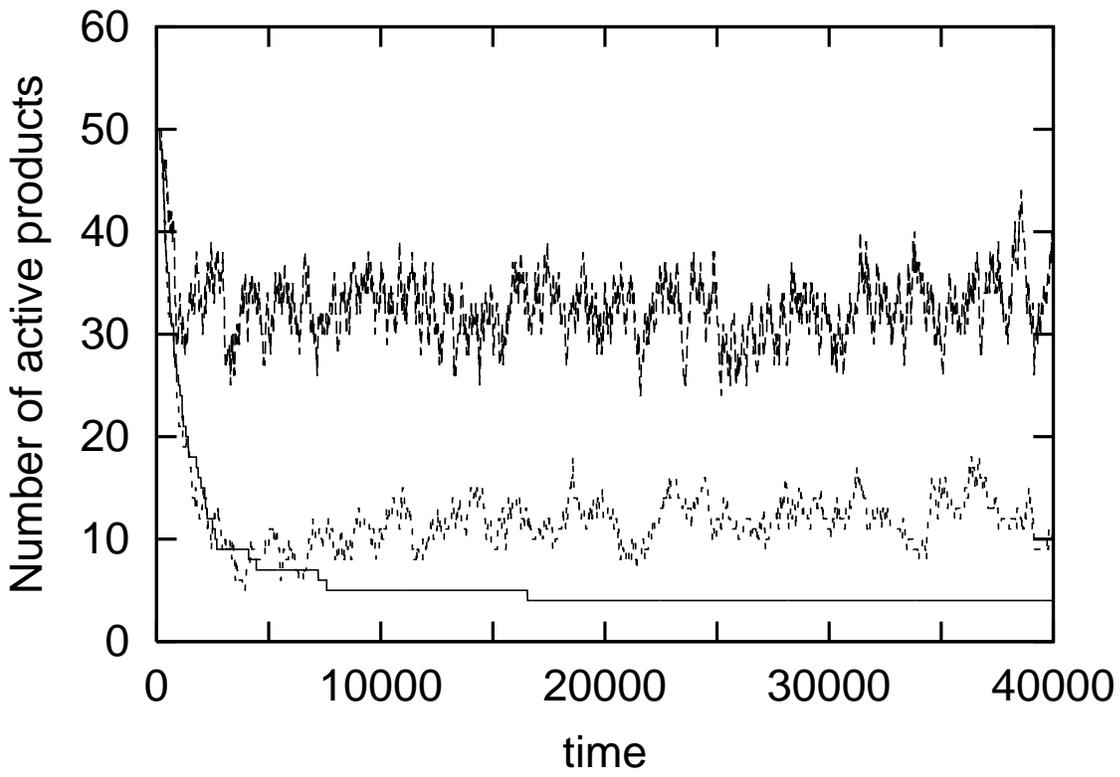
\item Fig. 2. Number of products that are not in the absorbing state
as function of time, with time defined as number of trades per agent,
for the cases of no production (full line), production rate 
$p=0.001$ (short dashed line), and $p=0.01$ (long dashed line). 
See text for the system parameters and other details.

\item Fig. 3a. Time course of the demand $D(j)$ of a given product $j$,
and of its volatility $V(j,t)\propto \cdot |D(j,t)-D(j,t-1)|/D(j,t-1)$.
One observe that the demand has persistent fluctuations, and that the 
volatility $V(j)$ tends to cluster. The system parameters are the same
of Fig.~2.\\
Fig. 3b. Hurst analysis of fluctuations in demand. 
In this case we considered a larger system,
$N_{ag}=500$, $N_{pr}=500$, $N_{mem}=1000$ and $N_{unit}=1000$,
thus spanning a larger scaling regime than for the smaller systems
considered before.
We see that the variation in demand may be fitted with a power law
$\Delta D \propto (\Delta t)^H$, where $H$ is the Hurst exponent,
$H =0.65 \pm 0.05$.

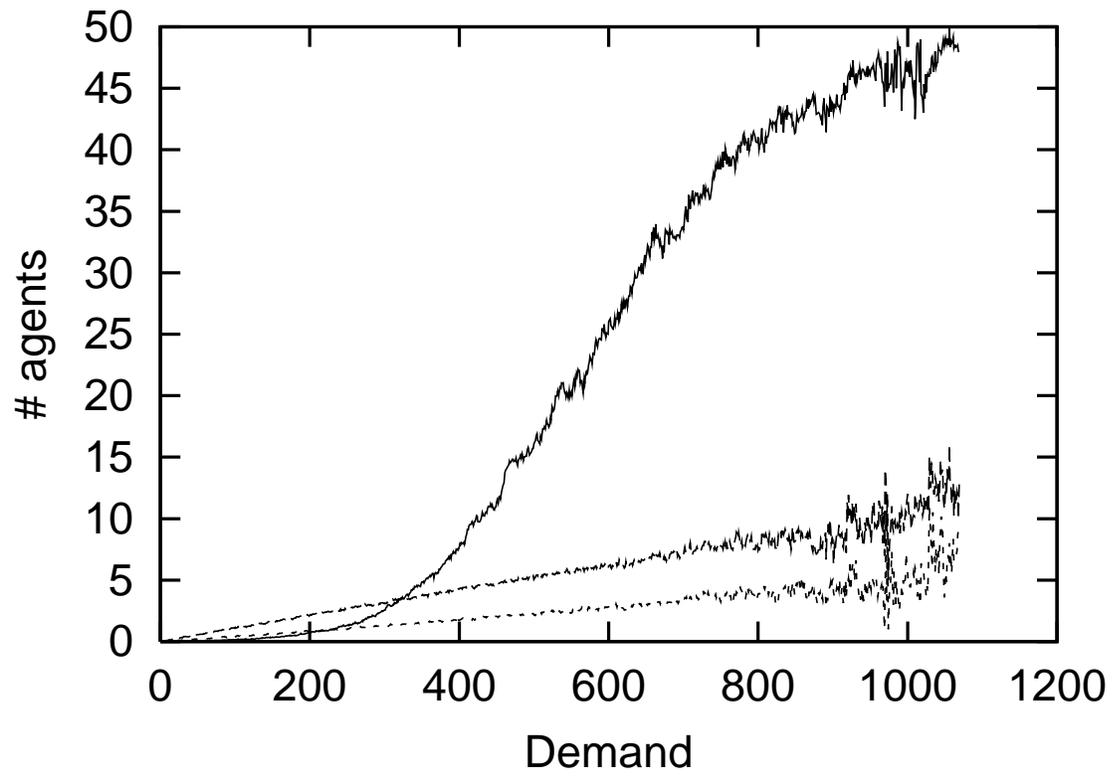
\item Fig.~4. Product properties as function of how much they are in demand.
The parameter values are the same of Fig.~2, with a production rate of
$p=0.001$.
The upper curve shows the monetary value of the product, as defined in the 
text, which shows a transition to global acceptance for demand values above 
$\approx 600$.
The lower curves show the number of agents that do not posess the product,
measured instantanously (middle curve), or over a time interval
where each product is traded in average two times (lower curve).
\end{itemize} 


\begin{thebibliography}{10}
\bibitem{DS00}
R. Donangelo and K. Sneppen, Physica A {\bf 276} (2000) 572.

\bibitem{Yasutomi}
A. Yasutomi, Physica D {\bf 82} (1995) 180.

\bibitem{Menger} 
C. Menger, {\it Principles of Economics\/} (Libertarian Press, Grove City, 
1994)

\bibitem{Hurst} H.E. Hurst, Transactions of the American Society of 
Civil Engineers, {\bf 116} (1951) 770.

\bibitem{f88} J. Feder, {\it Fractals\/} (Plenum Press, New Yourk, 1988).

\bibitem{Evertz}
C.J.G. Evertz, Fractals {\bf 3} (1995) 609.

\bibitem{RHSS}
R. Donangelo, A. Hansen, K. Sneppen and S. Souza,
Physica A, (2000) in press.

\bibitem{BPS}
P. Bak, M. Paczuski and M. Shubik, Physica A {\bf 246}, (1997) 430.

\bibitem{Farmer}
D. Farmer, Comp.\ in Science and Eng.\ {\bf 1}, (6), (1999) 26.


\end{thebibliography}
\end{document}